\documentclass{article}
\usepackage{ltwol2e}

\usepackage{epsfig}

\def\be{\begin{equation}}
\def\ee{\end{equation}}
\def\bea{\begin{eqnarray}}
\def\eea{\end{eqnarray}}

\begin{document}

\title{ANALYTIC QCD FLAVOR THRESHOLDS THROUGH TWO LOOPS} 

\author{MICHAEL MELLES}

\address{Department of Physics, Durham University, South 
Road, Durham DH1 3LE, England\\E-mail: Michael.Melles@durham.ac.uk}  


\twocolumn[\maketitle\abstracts{ 
In this talk \footnote{Presented at ICHEP'98, Vancouver, CA, July 1998}, 
we present a recently suggested way on how to
analytically incorporate massive threshold effects into observables
calculated in massless QCD. No matching is required since the renormalization
scale is in this approach connected to the physical momentum transfer
between static quarks in a color singlet state. We discuss
massive fermionic corrections to the heavy quark
potential through two loops. The calculation uses a mixed approach of
analytical, computer-algebraic and numerical tools including Monte
Carlo integration of finite
terms. Strong consistency checks are performed by
ensuring the proper cancellation of all non-local divergences by the
appropriate counterterms and by comparing with the massless limit.
The size of the effect for the (gauge invariant) fermionic part of
$\alpha_V ( {
\bf q^2},m^2 ) $
relative to the massless case at the charm and
bottom flavor thresholds is found to be of order $33 \%$.
}]

\section{Introduction} \label{sec:intro}

\subsection{Analytic Thresholds at one Loop}

In the $MS$ and $\overline{MS}$ renormalization schemes, the running
of the QCD coupling $\alpha_s$, by construction, does not ``know'' about the 
masses, $m_q$, of  
quarks. The $\beta$ function \cite{Callan,Symanzik} describes the evolution of the strong coupling
``constant'' in the asymptotic regime, i.e. for values of the renormalization scale
$\mu \gg m_q$. Near the quark flavor thresholds one has to turn to effective 
descriptions which match threories with $n$ massless quarks onto a theory with
$n-1$ massless and one massive flavor at the ``heavy'' quark threshold. In this way
the dependence on the dimensional regularization mass parameter $\mu$ is reduced to
next to leading order effects by giving up the analyticity of the coupling at the
flavor threshold \cite{Soper,Marciano,Santamaria}.

While this procedure of matching conditions and effective descriptions is
certainly workable, from a theoretical standpoint it would be advantageous
to have a physical coupling constant definition which is analytic at thresholds.
In addition, as a physical observable, the total derivative with respect to the
renormalization scale $\mu$ vanishes.
Such a system is given by identifying the ground state energy of the vacuum
expectation value of the Wilson loop as
the potential $V$ between a static quark-antiquark pair in a
color singlet state \cite{Susskind,Fischler,Dine,Feinberg,Billoire,Peter}:

\begin{equation}
V( r, m^2_q) = - \lim_{t \rightarrow \infty} \frac{1}{it} log \langle 0| Tr \; \{
P \; exp \left( \oint dx_\mu A^\mu_a T^a \right) \} |0 \rangle \label{eq:Vdef}
\end{equation}

\noindent
where $r$ denotes the relative distance between the heavy quarks, $m_q$ the mass
of ``light'' quarks contributing through loop effects and $T^a$ the generators
of the gauge group. It is then convenient to define the effective charge
$\alpha_V( {\bf q}^2,m^2)$ as

\begin{equation}
V({\bf q}^2,m_q^2) \equiv - \frac{4 \pi C_F \alpha_V ( {\bf q}^2,m_q^2)}{{\bf q}^2}
\label{eq:aVdef}
\end{equation}

\noindent
in momentum space with ${\bf q}^2 \equiv q_0^2-q^2=-q^2 > 0$. The factor $C_F$ is the value of the Casimir operator $T^aT^a
$
in the fundamental representation of the external sources and factors out to all
orders in perturbation theory. As one is free to choose the representation
of the external particles, we obtain the static gluino potential by adopting the
adjoint representation.

Recently \cite{Melles1l}, the effect of the massive fermionic one loop contributions to
the heavy quark potential were incorporated into a continuous and smooth function
$n_f \left( \rho_q \right)$ given to lowest order by

\begin{eqnarray}
n^o_{f,V} \left( \rho_q \right) &\equiv& \frac{3 \pi}{\alpha_V^2} \frac{ \partial \alpha_V^{
f_{1loop}}}{\partial Q} \nonumber \\
&=& 1-\frac{6}{\rho_q} + \frac{24}{\rho^\frac{3}{2}_q \sqrt{4+\rho_q}} \tanh^{-1} 
\sqrt{\frac{\rho_q}{\rho_q+4}} \label{eq:nf}
\end{eqnarray}

\noindent where $\rho_q \equiv \frac{Q^2}{m_q^2}$ and $Q^2 \equiv -q^2$. The mass dependence
of the physical $V$-scheme can now be transferred to the $\overline{MS}$ scheme
by using the commensurate scale relation \cite{csr} between the two schemes. 
Employing the multi scale approach of Ref. \cite{csr} gives the following scale fixed
relation through two loops \cite{Melles1l}:

\begin{equation}
\alpha_{\overline{MS}} (Q) = \alpha_V(Q^*)+2 \frac{\alpha^2_V(Q^{**})}{\pi}
+4.625 \frac{\alpha^3_V(Q^{**})}{\pi^2} \label{eq:csrMSV}
\end{equation}

\noindent with

\begin{equation}
Q^*=2.3Q \;\;,\;\;Q^{**}=6.539Q
\end{equation}

\noindent whereas $Q^{***}$ to this order is not constrained. A first approximation
is obtained, however, by setting $Q^{***}=Q^{**}$ \cite{csr}. Note that the scale
$Q$ at one loop is a factor $0.4$ smaller than the physical scale $Q^*$.

\subsection{Analytic ${\tilde \alpha}_{\overline{MS}}$}

One is now free to adopt the commensurate scale relation of Eq. \ref{eq:csrMSV}
as a definition of the extended scheme ${\tilde \alpha}_{\overline{MS}}$ \cite{Melles1l}.
At one loop we have therefore

\begin{equation}
{\tilde \alpha}_{\overline{MS}} (Q) \equiv \alpha_V(Q^*)+2 \frac{\alpha^2_V(Q^{**})}{\pi}
\label{eq:analams}
\end{equation}

\noindent
for all scales $Q$. Eq. \ref{eq:analams} not only provides an analytic extension of
$MS$ like renormalized schemes, but it also ties down the renormalization scale
$\mu$ to the physical scale with massive quarks, entering into the vacuum polarization
contributions to $\alpha_V$. There is thus no scale ambiguity in perturbative 
expansions in $\alpha_V$ or ${\tilde \alpha}_{\overline{MS}}$. To lowest order
we obtain in addition

\begin{equation}
{\tilde n}^o_{f,{\overline{MS}}} \left( \frac{Q^2}{m_q^2} \right) = n^o_{f,V}
\left( \frac{{Q^*}^2}{m_q^2} \right)
\end{equation}

\noindent
A very good ($\approx 1 \%$) approximation is given by the following simple result
for the one loop effective function of flavors:

\begin{equation}
{\tilde n}^o_{f,{\overline{MS}}} \left( \frac{Q^2}{m_q^2} \right) \approx
\frac{1}{1+\frac{5}{\rho_q}} \label{eq:analnf}
\end{equation}

\begin{figure}
\center
\epsfig{file=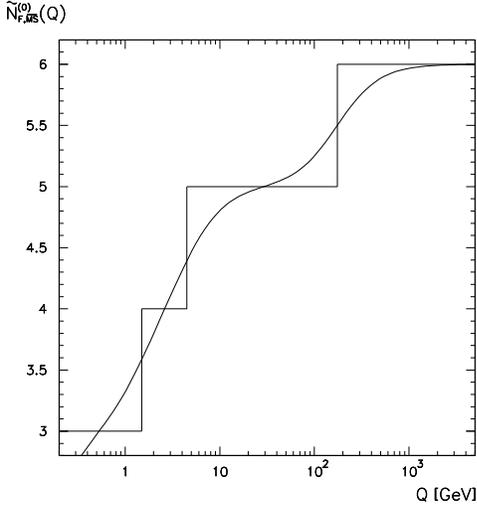,height=2.8in}
\caption{The sum over massive quark flavors for the effective analytic function
${\tilde n}^o_{f,{\overline{MS}}} \left( \frac{Q^2}{m_q^2} \right)$ given in Eq.
\ref{eq:analnf}.}
\label{fig:analnf}
\end{figure}

\noindent
Fig. \ref{fig:analnf} contains the analytic function $n_f$ summed over all quark
flavors in comparison with the conventional step-function approach, which 
treats quarks as infinitely heavy below threshold and massless above.

\subsection{Applications}

It is of course possible to treat mass effects exactly within the 
$\overline{MS}$ scheme \cite{Chetyrkin,Hoang} by explicitly calculating higher twist QCD 
corrections.
In order to have a meaningful comparison between the different approaches of
incorporating mass effects, we choose to compare the above treatment to the
explicit corrections to an observable, here the quark part of the non-singlet
hadronic width of the Z-boson, $\Gamma^{NS}_{had,q}$. Writing the QCD corrections
in terms of an effective charge we have:

\begin{equation}
\Gamma_{had,q}^{NS}=\frac{G_FM_Z^3}{2\pi\sqrt{2}}
\sum_{q}\{(g_V^{q})^2+(g_A^{q})^2\}
\left[1+\frac{3}{4}C_F\frac{\alpha_{\Gamma,q}^{NS}(s)}{\pi}\right]
\end{equation}

\noindent
where the effective charge $\alpha_{\Gamma,q}^{NS}(s)$ contains all
perturbative QCD corrections. At the one loop level we are left with a very simple
expression

\begin{eqnarray}
\label{eq:aganalytic}
\frac{\alpha_{\Gamma,q}^{NS}(s)}{\pi} & = &
\frac{{\tilde \alpha}_{\overline{MS}}(Q^*)}{\pi}.
\end{eqnarray}

\begin{figure}
\center
\epsfig{file=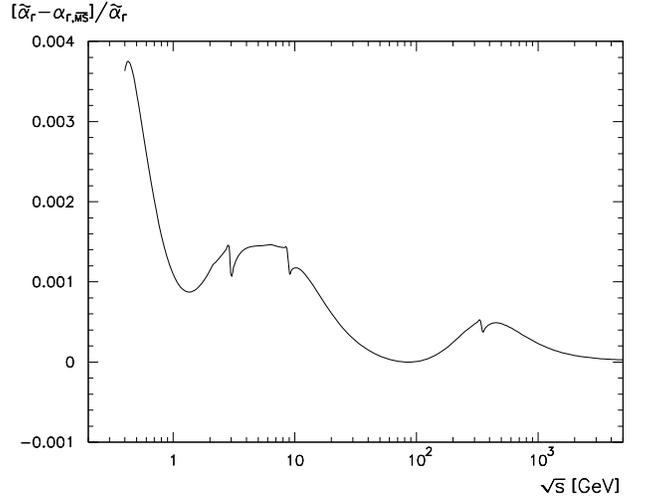,height=2.8in}
\caption[*]{The relative difference between the calculation of
$\alpha_{\Gamma,q}^{NS}(s)$ in the analytic extension of the $\overline{MS}$ scheme
and the standard treatment of masses in the $\overline{MS}$ scheme. 
The discontinuities
are due to the mismatch between the $s/m^2$ and $m^2/s$ expansions of the
explicit QCD corrections in Refs. \cite{Chetyrkin,Hoang}.}
\label{fig:GZcomp}
\end{figure}

\noindent
This simple expression reflects the fact that the effects of quarks in the
perturbative coefficients, both massless and massive, should be absorbed
into the running of the coupling. The BLM-scale \cite{blm} is 
$Q^*=0.7076 \sqrt{s}$ for this observable
\cite{Melles1l}. The explicit higher twist corrections were calculated in  Ref. \cite{Chetyrkin,Hoang} 
as expansions in $\frac{m_q^2}{s}$ and $\frac{s}{m_q^2}$. Fig. \ref{fig:GZcomp}
contains relative difference between the two approaches and can be seen to
be in about permille agreement for perturbative values of the energy.

This remarkable level of agreement implies that in order to incorporate 
one loop massive QCD-flavor threshold effects into 
observables which are known only in massless
QCD in the $\overline{MS}$ scheme, one just has to apply the analogous 
steps as above, namely replacing

\begin{equation}
n_f \longrightarrow {\tilde n}^o_{f,{\overline{MS}}} \left( \frac{Q^2}{m_q^2} 
\right) = n^o_{f,V} \left( \frac{{Q^*}^2}{m_q^2} \right)
\label{eq:analsub}
\end{equation}

\noindent where $Q^*$ is the BLM-scale of that process. In order to include
all flavors one simply
has the sum over all $m_q$ contributions as indicated in Fig. \ref{fig:analnf}.

\section{Two Loop Results} \label{sec:tl}

\begin{figure}
\center
\epsfig{file=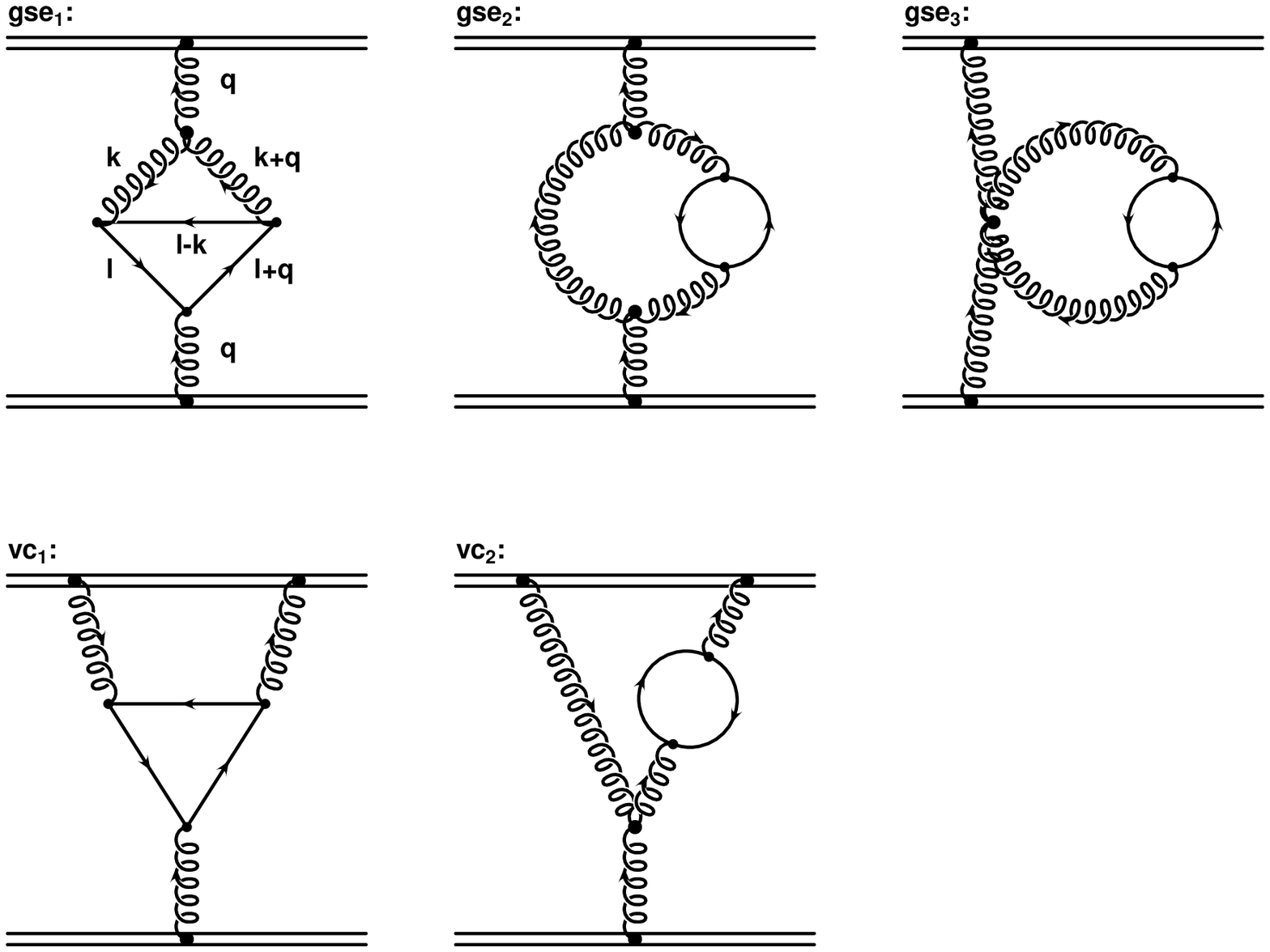,height=2.5in}
\vspace{0.5cm} \\
\epsfig{file=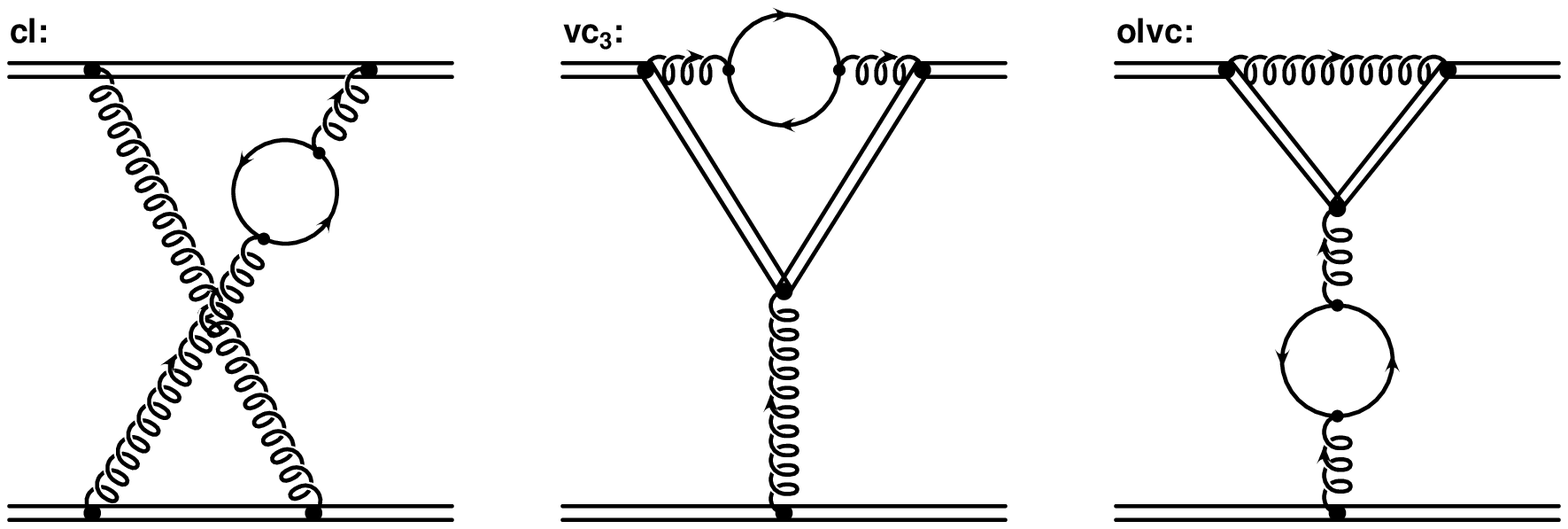,height=1.1in}
\vspace{0.5cm} \\
\epsfig{file=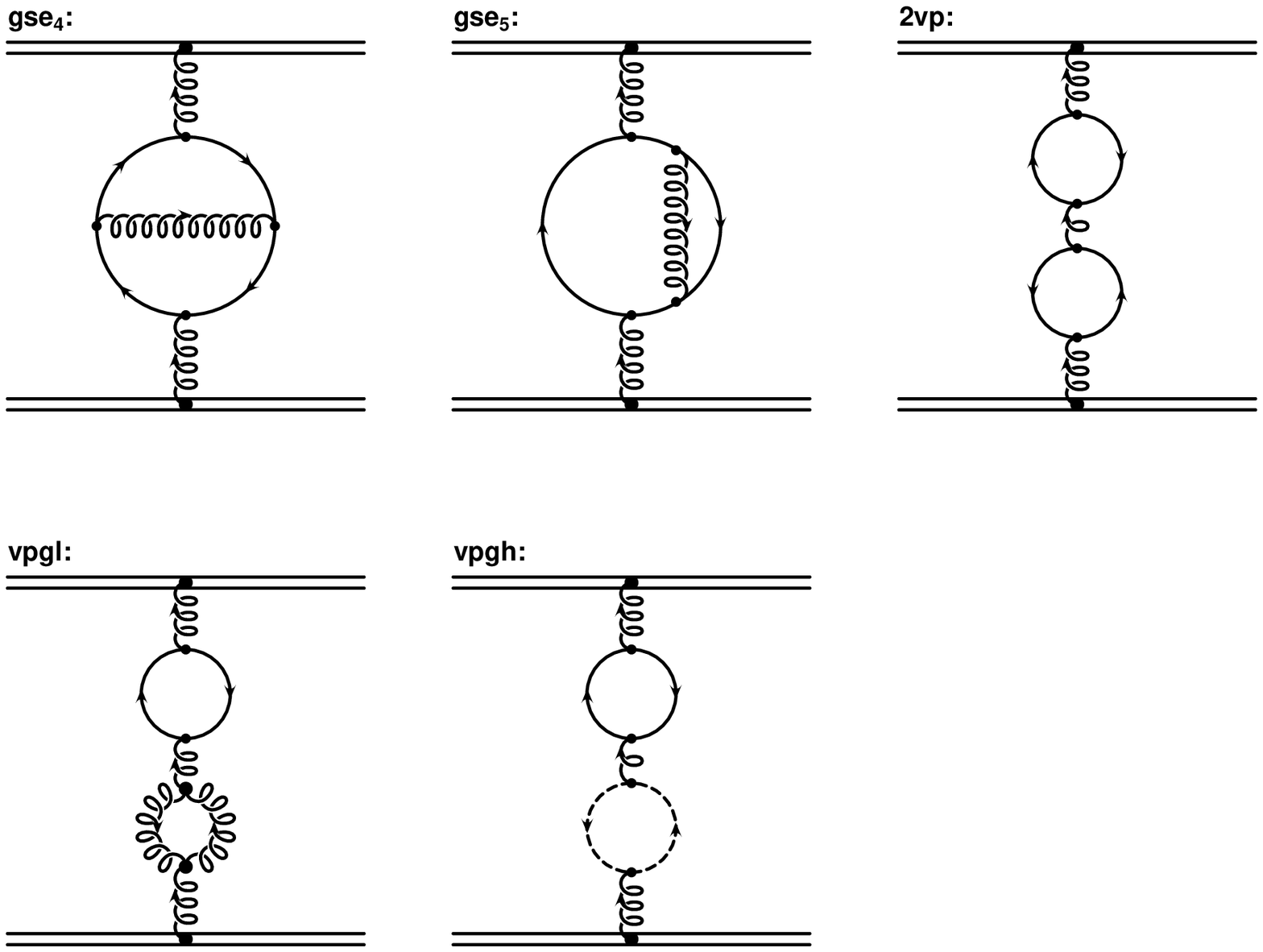,height=2.5in}
\caption{The two loop massive fermionic corrections to the heavy quark potential.
The first two rows contain diagrams with a typical non-Abelian topology.
Double lines denote the heavy quarks, single lines the ``light'' quarks. Color
and Lorentz indices are suppressed in the first graph. The notation for
the remaining digrams is analogous.
The middle line includes the infra-red divergent ``Abelian'' Feynman diagrams.
While the topology of these three diagrams is the same as in QED, they
contribute to the potential only in the non-Abelian theory due to
color factors ~$C_F C_A$. In addition, although each diagram is infra-red
divergent, their sum is infra-red finite.
The infra-red finite Feynman diagrams with an Abelian topology
are shown in the last two rows 
plus diagrams consisting of one loop insertions with non-Abelian terms.}
\label{fig:tlfd}
\end{figure}

At the two loop level the situation becomes much more cumbersome. The massless
case including pure gluonic corrections was calculated in Ref. \cite{Peter}. Recently, the
massive two loop corrections, depicted in Fig. \ref{fig:tlfd}, were obtained
in Ref. \cite{Melles2l}.

The double lines denote the static color sources for which one employs the
heavy quark effective Feynman rules (HQET), see for example Refs. \cite{Neubert}, while
the rest contains the full QCD dynamics including massive fermion lines.
The results in Ref. \cite{Melles2l} were obtained in the $MS$ renormalization scheme,
related to the $\overline{MS}$ scheme through a simple scale shift

\begin{equation}
\mu_{_{MS}}= \sqrt{\frac{e^\gamma}{4 \pi}} \mu_{_{\overline{MS}}} \;\;,
\end{equation}

\noindent by a combined analytical and numerical approach. For the two point
functions a tensor decomposition was performed following the techniques 
given in Ref. \cite{Weiglein}. The algebraic manipulation language FORM \cite{form}
was used for this purpose. The resulting scalar integrals are then solved by
calculating the pole terms analytically and integrating over the remaining
Feynman parameters numerically. 

For the higher point functions this approach is no longer applicable due to 
the presence of the heavy quark propagators. It was found to be advantageous
to integrate out the fermion loop analytically and then proceed with the
remaining integrations. Details are given in Ref. \cite{Melles2l}.

The required expansions in powers of $\epsilon
=4-n$ were performed by MAPLE as was the translation of finite parts into
FORTRAN. This latter point is crucial in face of the enormous complexity of the
obtained results.

\begin{figure}
\center
\epsfig{file=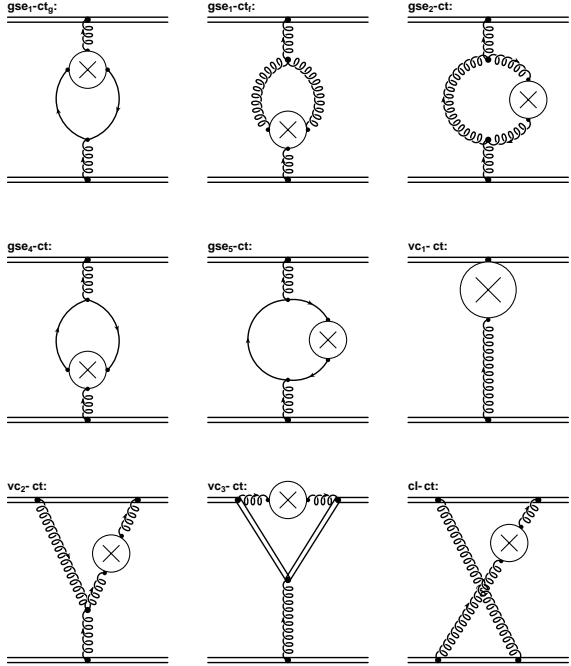,height=3.5in}
\caption{The counterterms corresponding to the massive two loop diagrams of 
Fig. \ref{fig:tlfd}.}
\label{fig:tlct}
\end{figure}

Strong consistency
checks were performed including the explicit calculation of all $MS$ counterterms,
shown in Fig. \ref{fig:tlct},
to ensure the locality of renormalization constants, the correct gluon wave function
renormalization constant and agreement with the $\frac{1}{\epsilon^2}$, 
$\frac{1}{\epsilon}$ pole terms from the massless calculation for the 
remaining three and four point two loop functions.

\begin{figure}
\center
\epsfig{file=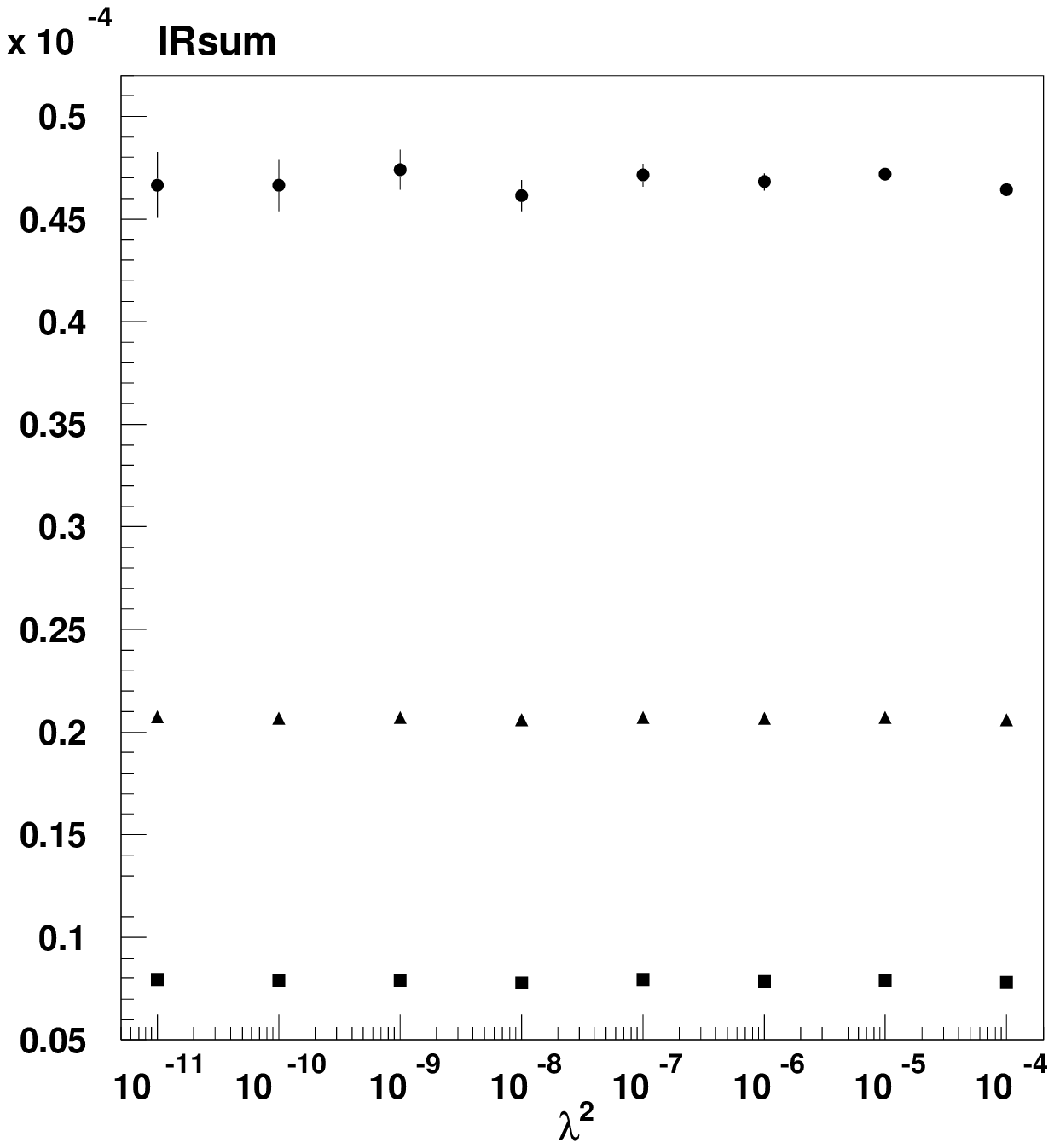,height=3.0in}
\caption{The sum of the $\lambda^2$-dependent amplitudes
and counterterms
${\cal M}^{k_0}_{cl}+{\cal M}^{k_0}_{vc_3}+{\cal M}_{olvc}+
{\cal M}^{k_0}_{cl_{ct}}+{\cal M}^{k_0}_{{vc_3}_{ct}}$. Circles
correspond to a choice of $q^2 = -10 GeV^2$ and $m = m_c$,
triangles to $q^2 = -100 GeV^2$ and
$m = m_c$ while the lower curve (squares) has $q^2 = -100 GeV^2$ and $m = m_b$.
the overall normalization neglects color factors and the coupling
strength. All data are obtained by using $10^6$ evaluations per iteration
with VEGAS and 100 iterations. The statistical error is
indicated and smaller than the symbols where invisible.
The sum for each of the displayed sets of parameters
is clearly independent of the IR-gluon mass regulator $\lambda$ as expected.}
\label{fig:lplot}
\end{figure}

\noindent
Furthermore, the absence of infrared divergences in the sum of the
$MS$ renormalized diagrams 
${\cal M}_{cl}$, ${\cal M}_{{vc}_3}$ and ${\cal M}_{olvc}$, 
which all contain infrared poles individually, 
is demonstrated in Fig. \ref{fig:lplot}. $\frac{i}{k_0}$ denotes the heavy quark
propagator and only terms with a dependence on $k_0$ needed to be regulated with
a gluon mass. The results clearly demonstrate the infrared finiteness of the
sum which contributes to the physical potential in Eq. \ref{eq:Vdef}.

The reason why these three amplitudes with an Abelian topology do contribute to 
the QCD potential but
are absent in QED is connected to the exponentiation that is implicit in Eq.
\ref{eq:Vdef}. The terms proportional to $C_F^2$ are actually already included
in the potential by the exponentiation of the lower order Born and vacuum 
polarization contributions. In the non-Abelian theory, however, there is also
a contribution proportional to $C_FC_A$, which cannot be obtained by the lower
order terms and thus must be taken into account for the static QCD potential.

\begin{figure}
\center
\epsfig{file=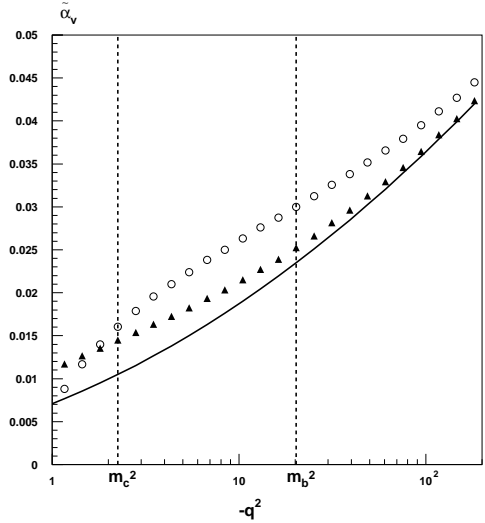,height=3.0in}
\caption{The complete two loop mass dependence of $\tilde{\alpha_V} \equiv \frac{
\alpha^{f_{2\,loop}}_V}{g^6}$ for $m^2=m_c^2=(1.5 GeV)^2$ (triangles) and $m^2=m_b^2
=(4.5 GeV)^2$ (open circles). The massless case is also included (line).
In all three curves we use $\mu=0.031$. The deviation from the massless case
at the flavor thresholds is of order of $33\%$ and is dominated by the new
non-Abelian contributions}
\label{fig:aV}
\end{figure}

We also found perfect agreement
for the numerical results for the finite, $MS$ renormalized
expressions in the limit $m_q \longrightarrow 0$.
The finite expressions were calculated with the Monte Carlo integrator VEGAS \cite{vegas}
and Fig. \ref{fig:aV} displays the weighted sum of all two loop 
$MS$ renormalized massive fermionic
corrections to $\alpha_V$.
 
It was found that the overall curve is
dominated by the non-Abelian threshold behavior (partially due to the extra factor
of $C_A$). The ``$m_c$-graph'' (triangles) matches the massless case for lower
values of $-q^2$ as $m_c^2 \ll m_b^2$. At the respective thresholds we find
roughly a 33 \% deviation relative to the massless case. This could be very
significant for applications where quark masses are expected to play
an important part.
At high values of $-q^2$ the theory becomes massless and reproduces
the leading logarithmic terms obtained by the $\beta$-function analysis as
these coefficients are scheme independent through two loops in a massless
theory.
These analyses can also be helpful for the incorporation of
massive fermions in lattice analyses as the heavy quark potential is defined
by the gauge invariant vacuum expectation value of the Wilson loop in Eq. 
\ref{eq:Vdef}.

\section{Conclusions}

There is a very elegant and simple way
to describe analytically massive one loop QCD flavor threshold effects in 
observables calculated
in massless QCD in the $\overline{MS}$ renormalization scheme. All one needs to
do is to substitute the discontinuous function of active flavors, $n_f$, according
to Eq. \ref{eq:analsub}. There is thus no need for complicated higher twist
calculations in the $\overline{MS}$ scheme. 
At two loops, the level of perturbation theory 
to which many 
observables are known in massless (!) QCD, 
it should be possible to extend the 
one loop analysis presented in section \ref{sec:intro} by using the now available
explicit two loop results discussed in section \ref{sec:tl}. 

The situation is now more complicated, however. The only feasible approach is
a numerical differentiation of the obtained Monte Carlo results. 
This implies questions relating to the technical precision domain.
In addition one
needs to include running mass effects at the one loop level as they enter into 
the two loop analysis.
Work towards this end is in progress \cite{Melles} and will hopefully 
soon lead to a two loop
extension of an effective analytic function of quark flavors.

\section*{Acknowledgements}
The author would like to thank S.J. Brodsky, M. Gill and J. Rathsman for their
contributions to the one loop results. This work was supported by
Deutsche Forschungsgemeinschaft, Reference \# Me 1543/1-1
and the European Union TMR-fund.

\end{document}